\documentclass[copyright]{eptcs}
\providecommand{\event}{DCFS 2010} 

\usepackage{breakurl}             
\usepackage{amsmath}
\usepackage{amsfonts}
\usepackage{amssymb}
\usepackage{graphicx}
\usepackage{verbatim}

\newtheorem{theorem}{Theorem}
\newtheorem{lemma}{Lemma}

\newcommand{\eps}{\varepsilon}

\newcommand{\nsc}{\mathrm{nsc}}

\renewcommand{\sc}{\mathrm{sc}}
\renewcommand{\le}{\leqslant}
\renewcommand{\ge}{\geqslant}

\title{Complexity in Prefix-Free Regular Languages}
\author{Galina Jir\'askov\'a
\thanks{Research supported by VEGA grant 2/0111/09}
\institute{Mathematical Institute\\ Slovak Academy of Sciences\\
   Gre{\v s}{\' a}kova 6, 040 01 Ko\v{s}ice\\ Slovak Republic}
\email{jiraskov@saske.sk}
\and
 Monika Krausov\'a
\institute{Institute of Computer Science\\ Pavol Jozef \v Saf\'arik University,\\
   Jesenn\'a 5, 041 54 Ko\v{s}ice\\ Slovak Republic}
\email{mon.krausova@gmail.com}
}
\def\titlerunning{Complexity in Prefix-Free Regular Languages}
\def\authorrunning{G. Jir\'askov\'a \& M. Krausov\'a}

\begin{document}
\maketitle

\begin{abstract}
  We examine  deterministic and nondeterministic state complexities
  of regular operations on prefix-free languages.
  We strengthen several results by providing witness languages
  over smaller alphabets, usually as small as possible.
  We next provide the tight bounds
  on  state complexity of symmetric difference,
  and deterministic and nondeterministic state complexity of
  difference and cyclic shift of prefix-free languages.
\end{abstract}

\section{Introduction}
\label{***intro}
 A language  is prefix-free 
 if for every string  in  the language,
 no proper prefix of the string is in the language.
 Deterministic and nondeterministic state complexity 
 of basic  operations on prefix-free regular languages
 have  recently been studied by Han and Salomaa \cite{hs09n,hs09}.
 The two papers follow   current research that focuses 
 on complexity in  various sublasses of regular languages
 \cite{bhk09,bjl10,bjz10,hs09s}.

 Here we continue this research and study
 the descriptional complexity of regular operations
 in the class of prefix-free regular languages.
 We strengthen several results on state complexity
 in \cite{hs09n,hs09}
 by providing witness languages over smaller alphabets,
 usually as small as possible.
 We also correct some errors in these two papers,
 in particular, the binary automata used for the result on reversal
 do not provide the claimed lower bound.
 We next study the state complexity of
 difference, symmetric difference, and cyclic shift,
 and provide  tight bounds.

 In the second part of the paper,
 we examine the nondeterministic state complexity of regular operations.
 We introduce a new fooling-set lemma,
 which allows us to give a correct proof for union,
 and to get the tight bound for cyclic shift.
 The idea behind the lemma is
 to find a fooling-set for a regular language
 and then show  that one more state is necessary
 by finding two appropriate strings. 
 We prove tight bounds on the nondeterministic state complexity
 of all basic operations including difference and cyclic shift.

 \section{State Complexity in Prefix-Free Languages}
 \label{***state}

 We start with investigation of
 state complexity of regular operations on prefix-free languages.
 The languages are represented by minimal dfa's,
 thus each of the dfa's has
 exactly one final state going to the dead state
 on every input symbol~\cite{hs09}.
 Then an operation is applied, and we are asking how many states,
 depending on the state complexities of operands,
 are sufficient and necessary in the worst case for a dfa
 to accept the language resulting from the operation.
 The next theorem provides the tight bounds for Boolean operations.
 In the case of union and intersection, the upper bounds are from \cite{hs09},
 where witness languages were defined 
 over a three- and four-letter alphabet, respectively.
 We provide binary witnesses for both operations.
 Then we study symmetric difference and difference,
 and get the tight bounds in the binary and ternary case, respectively.

 \begin{theorem}[Boolean Operations]\label{thm:sc_boolean}
  Let $m,n\ge 3$ and let  $K$ and $L$
  be  prefix-free regular languages  with $\sc(K)=m$ and $\sc(L)=n$.
  Then \\
  1. $\sc(K\cap L)\le mn-2(m+n)+6$, and the bound is tight in the binary case; \\
  2. $\sc(K\cup L)\le mn-2$, and the bound is tight in the binary case;\\
  3. $\sc(K\oplus L)\le mn-2$, and the bound is tight in the binary case;\\
  4. $\sc(K\setminus L)\le mn-m-2n+4$, and the bound is tight in the ternary case.
 \end{theorem}

 \noindent\emph{Proof.}
  Let the dfa's have states $0,1,\ldots,m-1$ and $0,1,\ldots,n-1$,
  of which $m-2$ and $n-2$ are final, and $m-1$ and $n-1$ are dead. 
  The initial state is 0.
  
\begin{figure}[t]
  \centerline{\includegraphics[scale=.39]{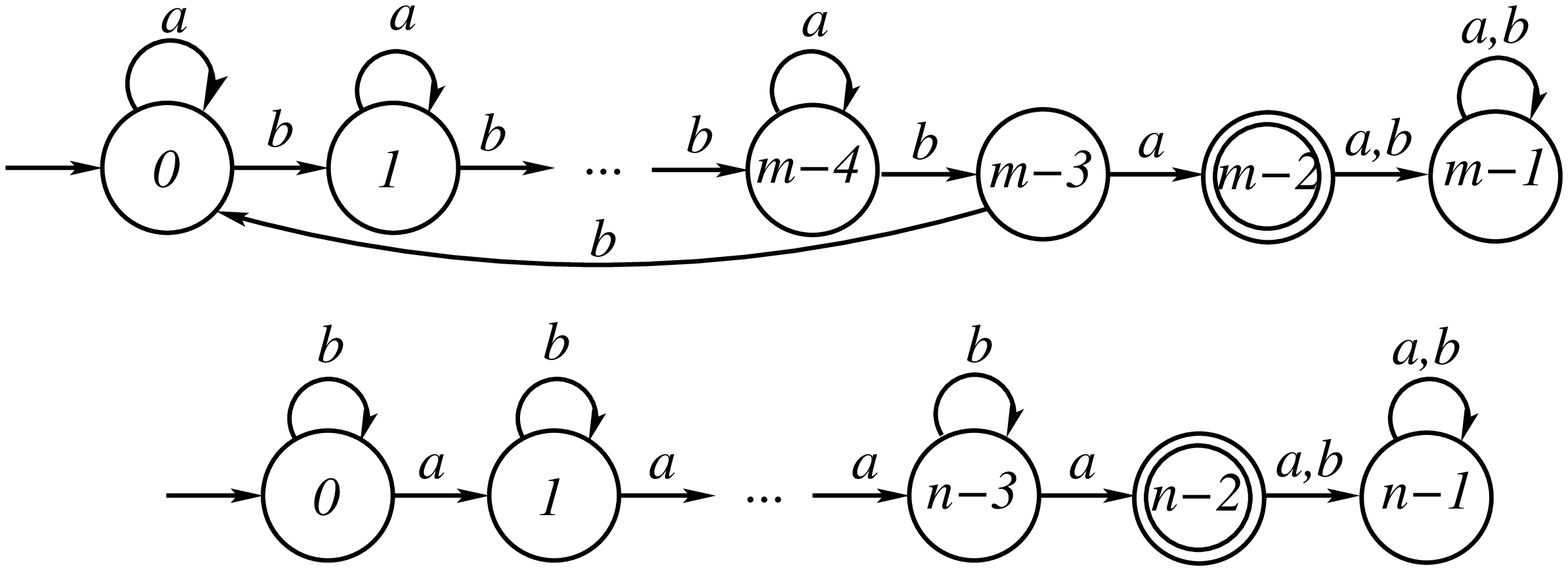}}
  \caption{The prefix-free  dfa's
           meeting the  bound $mn-2(m+n)+6$ for intersection.}
  \label{fig:sc_intersection}
  \end{figure}

  1.~For tightness, consider binary prefix-free 
  dfa's of Figure~\ref{fig:sc_intersection}.
  The strings
  $a^{n-3}b^{m-3}a$, $a^{n-3}b^{m-3}aa$, and $a^jb^i$ with $0\le i \le m-3$ 
  and $0 \le j \le n-3$ are pairwise distinct
  in the right-invariant congruence defined by language $K\cap L$.

  2.~Let 
  $K=(a^*b)^{m-2}$ and
  $L=(b^*a)^{n-2}$. 
  The strings  $b^ia^j$ with  $0\le i \le m-3$ and $0\le j \le n-1$,
  $a^jb^{m-2}$ and  $a^jb^{m-1}$ with $  0\le j \le n-3$, 
  and $a^{n-3}b^{m-2}a$ and $a^{n-3}b^{m-2}aa$
  are pairwise distinct 
  for  $K\cup L$.
 
  3.~In the cross-product automaton for symmetric difference, the rejecting 
  state $(m-2,n-2)$ is equivalent to the dead state, and 
  states $(m-2,n-1)$ and $(m-1,n-2)$ accept only $\varepsilon$.
  The same languages as for union meet the bound.

  4.~ All the states 
  of the cross-product automaton
  in the last row and state $(m-2,n-2)$
  are dead,  the other states in the last but one row only accept $\eps$.
  Pairs $(i,n-2)$ and $(i,n-1)$ are equivalent as well.
  This gives the upper bound, 
  which is met by
  $K=(b^*(a+c))^{m-2}$ and
  $L=((a+c)^*b)^{n-3}c^*(a+b)$.
  \hfill$\Box$\bigskip

 We now continue with concatenation and star,
 and slightly improve the results from \cite{hs09}
 by providing unary witnesses for concatenation,
 and the complexity of star in the unary case.

 \begin{theorem}[Concatenation and Star]\label{thm:con,star}
  Let $m,n\ge2$
  and let  $K$ and $L$ be  prefix-free regular languages with  $\sc(K)=m$. $\sc(L)=n$.
  Then \\
  1. $\sc(KL)\le m+n-2$ and the bound is tight in the unary case;\\
  2. $\sc(L^*)\le n$. The bound is tight in the binary case if $n\neq3$.\\
  The tight bound for star in the unary case is $n-2$ if $n\ge3$.
 \end{theorem}

 \noindent\emph{Proof.}
  1. We can get a dfa for the concatenation
  from the dfa's   as follows \cite{hs09}.
  We remove the dead state from the first dfa,
  and merge  the final state in the first dfa  
  with the initial state in the second dfa. 
  All transitions going from a non-final state
  in the first dfa to the dead state will go to the dead state in the second dfa.
  The resulting automaton is a dfa of $m+n-2$ states for  concatenation.
  The bound is met 
  by unary prefix-free languages $a^{m-2}$ and $a^{n-2}$.

  2. We make the final state initial,
  and redirect  transitions  from the final state
  to such states, to which they go from the start state.
  The resulting dfa for star has at most $n$ states.
  The upper bound  is met by the binary prefix-free  
  language $(a^{n-2})^*b$ \cite{hs09}.
  In the unary case, if $n\ge3$, the only $n$-state dfa 
  prefix-free language is $a^{n-2}$.
  The star of this language, $(a^{n-2})^*$, is an $(n-2)$-state dfa language.
 \hfill$\Box$\medskip

 Before dealing with reversal, let us investigate  nfa-to-dfa conversion.
 We recall the result from \cite[Theorem 19, which uses the proof of Theorem 6,
 which in turn uses Moore's proof in \cite{mo71}]{bhk09}.
 We present different ternary witnesses, and give a simple proof. 
 Then we show that the  bound cannot be met in the binary case.

 \begin{theorem}[NFA to DFA Conversion]\label{thm:nfa-to-dfa}
  Let $n\ge 3$ and let $L$ be a prefix-free language  with $\nsc(L)=n$.
  Then $\sc(L)\le 2^{n-1}+1$.
  The bound is tight in the ternary case,  but cannot be met in the binary case.
 \end{theorem}

 \noindent\emph{Proof.}
  Consider an $n$-state nfa recognizing a non-empty prefix-free language.
  The corresponding minimal dfa has
  exactly one final state,
  and so we can merge all final states in the subset automaton.
  This gives the upper bound $2^{n-1}+1$.

\begin{figure}[b]
  \centerline{\includegraphics[scale=.40]{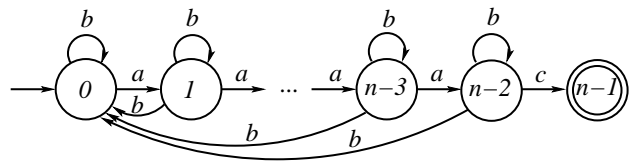}}
  \caption{The prefix-free  nfa
            meeting the  bound $2^{n-1}+1$ for nfa-to-dfa conversion.}
  \label{fig:nfa-to-dfa}
  \end{figure}

  For tightness, consider the ternary nfa of Figure~\ref{fig:nfa-to-dfa}.
  In the corresponding subset automaton,
  each singleton set and the empty set are reachable.
  Each set $\{i_1,i_2,\ldots,i_k\}$ with $0\le i_1 <i_2< \cdots < i_k\le n-2$
  of size $k$   is reached from 
  set $\{i_2-i_1,i_3-i_1,\ldots,i_k-i_1\}$ of size $k-1$ by $ba^{i_1}$.
  Since for each state $i$, the string $a^{n-2-i}c$
  is accepted by the nfa only from state $i$,
  no two different states of the subset automaton are equivalent.

  Now consider the binary case.
  In a minimal binary $n$-state prefix-free nfa
  denote by  $n$ the final state,
  and by $n-1$ a state that goes to $n$ by a symbol~$a$.
  In the corresponding subset automaton,
  there must be a state $i$ in $\{1,2,\ldots,n-1\}$
  that goes to a non-empty subset $S$ of $\{1,2,\ldots,n-1\}$  
  by symbol $a$ because otherwise
  the nfa on states $\{1,2,\ldots,n-1\}$ would be unary, and so
  the number of reachable states in the corresponding subset automaton
  could not be $2^{n-1}$.
  Since all subsets of $\{1,2,\ldots,n-1\}$
  must be reachable, the subset $\{i,n-1\}$ is reachable.
  However,  subset $\{i,n-1\}$ goes to a superset of state $S\cup\{n\}$ by $a$,
  which in turn goes by a non-empty string  
  to an accepting state that is reached from the superset.
  This contradicts to prefix-freeness of the accepted language.
 \hfill$\Box$\medskip

  In the case of reversal, the result in \cite{hs09}
  uses  binary dfa's from \cite{swy04}.
  It is claimed in \cite[Theorem~3]{swy04} that 
  the automata meet the upper bound $2^n$
  on the state complexity of reversal. However, this is not  true.
  In the case of $n=8$, with initial and final state 1,
  the number of reachable states in the subset automaton 
  corresponding to the reverse of
  the dfa is 252 instead of 256: subsets 
  $\{1,4,5,8\}$, $ \{2,5,6,1\}$, $\{3,6,7,2\}$, and $\{4,7,8,3\}$
  cannot be reached from any subset by $b$ 
  since each of them  contains exactly one of states 1 and 3; 
  and by $a$, there is a cycle among these states.
  A similar reasoning shows that, whenever $n=8+4k$,
  the automata with the initial and final state 1 in \cite{swy04}
  do not meet the bound $2^n$.
  The  binary automata with a single accepting state 
  meeting the upper bound for reversal
  have recently been presented in \cite{se10}.
  We use them to get correct ternary prefix-free witnesses for reversal.

 \begin{theorem}[Reversal]\label{thm:sc_reversal}
  Let $n\ge4$ and let $L$ be a prefix-free regular language with $\sc(L)=n$.
  Then $\sc(L^R)\le 2^{n-2}+1$. The bound is tight in the ternary case,
  but cannot be met in the binary case. 
 \end{theorem}

 \noindent\emph{Proof.}
  We first construct an nfa for the reversal from the given dfa
  by removing the dead state,
  reversing all transitions,
  and switching the role of the initial and final state.
  Since no transition in the resulting nfa goes to the initial state,
  the corresponding subset automaton has at most $2^{n-2}+1$  states.
  
  \begin{figure}[t]
  \centerline{\includegraphics[scale=.40]{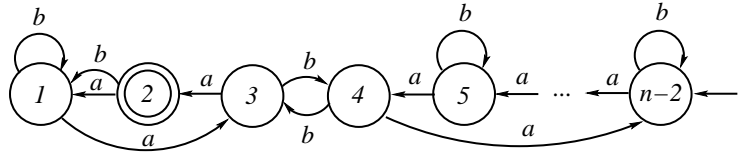}}
  \caption{The binary dfa requiring $2^{n-2}$ states for reversal.}
  \label{fig:binary}
  \end{figure} 

  For tightness, first consider  the binary dfa of $n-2$ states
  depicted in Figure~\ref{fig:binary}.
  It has been show in \cite{se10}, that the reversal of the language
  recognized by this dfa requires $2^{n-2}$ states.
  Now change the dfa as follows.
  Add two more states $n-1$ and $n$.
  State $n-1$ will be the sole final state, while state $n$ will be dead.
  Define transitions on a new symbol $c$:
  state 2 goes to the new final state $n-1$ by $c$,
  and each other state  goes to the dead state $n$.
  The resulting automaton is a prefix-free ternary $n$-state dfa 
  requiring $2^{n-2}+1$ deterministic states for reversal.
 
  Now consider the binary case. 
  Let $L$ be a binary prefix-free witness language.
  Then $\nsc(L^R)\le n-1$ because  the minimal dfa for $L$ has the dead state.
  Since $\sc(L^R)=2^{n-2}+1$,  language $L^R$
  is a binary witness for  nfa-to-dfa conversion.
  Theorem~\ref{thm:nfa-to-dfa} shows that this cannot happen.
 \hfill$\Box$\medskip

 The state complexity of cyclic shift was examined in \cite{jo08}, 
 where the upper and lower
 bound are only asymptotically tight. The next theorem provides the tight bound
 for this operation in the class of prefix-free regular languages.

 \begin{theorem}[Cyclic Shift]\label{thm:sc_cyclic}
  Let $L$ be a prefix-free  language with $\sc(L)=n$.
  Then $\sc(L^{cs})\le (2n-3)^{n-2}$. The bound is tight
  for a six-letter alphabet. 
 \end{theorem}

 \noindent\emph{Proof.}
  Consider an $n$-state dfa  for a prefix-free language $L$
  with states $1,2,\ldots, n$,
  of which 1 is the initial state,
  $n-1$ is the sole final state that goes to the dead state $n$ on each symbol.
  If a string $w$ is in the language $L^{cs}$,
  then $w=uv$ for some strings $u,v$ such that $vu\in L$.
  That is, the initial state 1 goes to a state $i$ by $v$,
  and then from state $i$ to the accepting state $n-1$ by $u$.
  Thus, a string $uv$ is in $L^{cs}$ if and only if
  there is a state $i$ such that $i$ goes to the accepting state $n-1$ by $u$,
  and the initial state 1 goes to state $i$ by $v$.
  Because of prefix-freeness,
  state $i$ is less then  $n-1$.
  Hence the cyclic shift is the union of $n-2$ concatenations
  $L(B_i)L(C_i)$, $i=1,2,\ldots,n-2$, where
  $B_i=(Q,\Sigma,\delta,i,\{n-1\} )$ and
  $C_i=(Q,\Sigma,\delta,1,\{i\} )$ (cf. \cite{jo08}).
  Each such concatenation is recognized by a dfa of $2n-3$ states
  since we first remove a dead state from $B_i$,
  then merge the final state of $B_i$ and the initial state of $C_i$,
  and finally merge states $n-1$ and $n$ in $C_i$
  since they are dead.
  Thus we have the union of $n-2$ dfa's,
  each of which has $2n-3$ states,
  which gives the upper bound $(2n-3)^{n-2}$.

\begin{figure}[t]
  \centerline{\includegraphics[scale=.40]{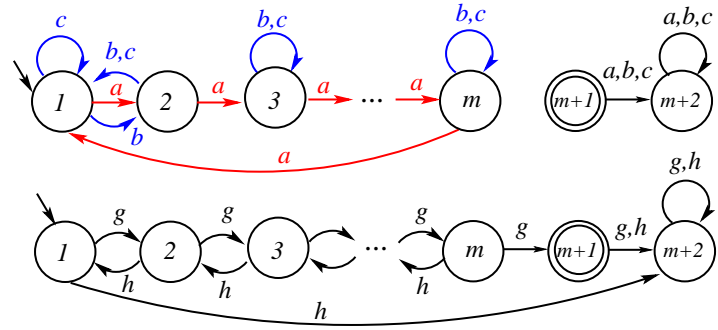}}
  \caption{The transitions on $a,b,c,g,h$ in the prefix-free 
           witness dfa for cyclic shift.}
  \label{fig:dfa_cyclic-dany}
  \end{figure}

  For tightness, set $m=n-2$ and let
  $\Sigma=\{a,b,c,d,g,h\}$.
  Define a prefix-free dfa  over $\Sigma$ of $n$ states $1,2,\ldots,m,m+1,m+2$,
  of which 1 is the initial state,
  $m+1$ is the sole accepting state that goes to the dead state $m+2$ 
  by each symbol;
  and for states $1,2,\ldots,m$, the transitions, except for symbol $d$, are defined
  as in Figure~\ref{fig:dfa_cyclic-dany}:
  Next,     by $d$, state $m$ goes to state $m+1$,
  and each other state to itself.
  The proof proceeds by showing the reachability 
  and inequivalence of all $m$-tuples in the subset automaton 
  corresponding to  $m(2m+1)$-state nfa for cyclic shift.
 \hfill$\Box$

 \section{Nondeterministic State Complexity}
 \label{***nondet}

 This section deals with the nondeterministic state complexity
 of regular operations on prefix-free languages.
 This time, the languages are represented by nfa's.
 The nfa's have exactly one final state that goes to the empty set by each symbol.
 However,  such an nfa is not guaranteed
 to accept a prefix-free language.
 On the other hand, if such an nfa is a partial dfa, 
 then it accepts a prefix-free language
 since to get the prefix-free dfa for the language 
 we only need to add a dead state.
 If accepted language consists of  strings ending in a symbol 
 that does not occur anywhere else in the string, 
 then such a language is prefix-free as well.

 We are asking how many states, depending 
 on the nondeterministic state complexity of operands, 
 are sufficient and necessary in the worst case for an nfa 
 with a single initial state
 to accept the language resulting from some operation.
 To prove the results we use
 a fooling set lower-bound technique. 
 A set of pairs of strings $\{(x_1,y_1),(x_2,y_2),\ldots,(x_n,y_n)\}$
 is called a \emph{fooling set} for a language $L$ if
 (1) for all $i$, the string $x_iy_i$ is in the language $L$, and
 (2) if $i\neq j$, then at least one of  strings
 $x_iy_j$ and $x_jy_i$ is not in the language $L$.
 It is well-known that the size of a fooling set for a regular language
 provides a lower bound on the number of states
 in any nfa for this language. 
 The next lemma shows that sometimes one more state is necessary.

 \begin{lemma}[\cite{jm10}]\label{le:fool}
  Let $L$ be a regular language.
  Let $\mathcal{A}$ and $\mathcal{B}$ be sets of pairs of strings
  and let $u$ and $v$ be two strings such that
  $\mathcal{A}\cup\mathcal{B}$,
  $\mathcal{A}\cup\{(\eps,u)\}$, and
  $\mathcal{B}\cup\{(\eps,v)\}$ are fooling sets for $L$.
  Then every nfa for $L$ has at least $|\mathcal{A}|+|\mathcal{B}|+1$ states.
 \hfill$\Box$
 \end{lemma}

 \begin{theorem}[Boolean Operations]\label{thm:boolean}
  Let $m,n\ge 3$. Let  $K$ and $L$
  be  prefix-free languages  with $\nsc(K)=m$ and $\nsc(L)=n$.
  Then \\
  1. $\nsc(K\cup L)\le m+n$, and the bound is tight in the binary case;\\
  2. $\nsc(K\cap L)\le mn-(m+n)+2$, and the bound is tight in the binary case;\\
  3. $\nsc(L^c)\le 2^{n-1}$,
     and the bound is tight in the ternary case;\\
  4. $\nsc(K\setminus L)\le (m-1)2^{n-1}+1$,
    and the bound is tight for a four-letter alphabet.
 \end{theorem}

 \noindent\emph{Proof.}
  1.~Let $A$ and $B$ be $m$ and $n$-state prefix-free nfa's
  with initial states $s_A$ and $s_B$,
  and transition functions $\delta_A$ and $\delta_B$, respectively.
  To get an nfa for the union we add a new initial state
  going to $\delta_A(s_A,a)\cup \delta_B(s_B,a)$ by each symbol $a$.
  Since both automata are prefix-free,
  we can merge their final states.
  Therefore, the upper bound is $m+n$.
  To prove tightness, consider  prefix-free languages
  $K=(a^{m-1})^*b$ and $L=(b^{n-1})^*a$
  accepted by an $m$-state and $n$-state nfa, respectively.
  Let $\mathcal{A}$ and $\mathcal{B}$ be the following set of pairs of strings:
  
  $\mathcal{A} = \{(a^{m-1}b,\eps)\}\cup\{(a^i,a^{m-1-i}b)
                       \mid i=1,2,\ldots, m-2\}\cup\{(a^{m-1},a^{m-1}b)\},$
                       
  $\mathcal{B}\  =  \{(b^j,b^{n-1-j}a)
                       \mid j=1,2,\ldots, n-2\}\cup\{(b^{n-1},b^{n-1}a)\}.$ \\
   Let us show that the set $\mathcal{A}\cup \mathcal{B}$
   is a fooling set for  language $K\cup L$.
   The concatenation of the first and the second part in each pair results
   in a string in $\{a^{m-1}b,a^{2m-2}b,b^{n-1}a,b^{2n-2}a\}$.
   Each of these strings is in language $K\cup L$.
   If we concatenate
   the first and the second part in two distinct pairs,
   we get a string in 
   $a^{m-1}b^+a^+(b+\eps)$ or in
   $\{a^rb, a^{m-1+r}b, b^sa, b^{n-1+s}a, \mid 0<r<m-1,0<s<n-1\}$
   or a string in $ a^+b^+a$.
   None of them  is in $K\cup L$.
   Next,  $\mathcal{A} \cup \{(\eps,b^{n-1}a)\}$
   and $\mathcal{B}\cup\{(\eps,a^{m-1}b)\}$
   are fooling sets for $K\cup L$.
   By Lemma~\ref{le:fool}, every nfa for the union
   has at least $m+n$ states. 
   Notice that the set of pairs in \cite{hs09n} is not a fooling set. 

\begin{figure}[tb]
  \centerline{\includegraphics[scale=.40]{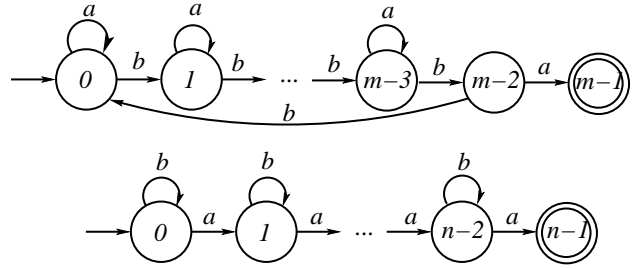}}
  \caption{The prefix-free  nfa's
           meeting the  bound $mn-(m+n)+2$ for intersection.}
  \label{fig:nsc_intersection}
  \end{figure}

\label{nsc_intersection}
  2.~In the cross-product automaton for the intersection,
  no string is accepted from states $(i,n-1)$ and $(m-1,j)$,
  except for the sole final state $(m-1,n-1)$.
  We can exclude all these states, and get an nfa of $(m-1)(n-1)+1$ states.
  For tightness, consider binary prefix-free
   nfa's of Figure~\ref{fig:nsc_intersection}.
  The languages are the same as in the deterministic case for intersection,
  but now they are accepted by nfa's, and so we do not need  dead states.
  Consider the cross-product nfa for the intersection of the two languages,
  and let $Q$ be the set of its $(m-1)(n-1)+1$ states excluding all states
  $(i,n-1)$ and $(m-1,j)$, but including state $(m-1,n-1)$.
   For each state $q$ in $Q$,
  there exist strings $x_q$ and $y_q$
  such that the initial state goes \emph{only} to state $q$ by $x_q$,
  and the string $y_q$ is accepted by the cross-product \emph{only} 
  from state $q$.
  It follows that  the set of pairs $\{(x_q,y_q) \mid q\in Q\}$
  is a fooling set for the intersection of the two languages
  since each string $x_qy_q$ is accepted by the cross-product automaton,
  while $x_py_q$ is not if $p\neq q$.

\label{part3}
  3. Let $N$ be an $n$-state  nfa for a prefix-free language $L$.
  The equivalent minimal dfa $D$ has exactly one final state $f$,
  from which all transitions go to the dead state $d$.
  It follows that  dfa $D$ has at most $2^{n-1}+1$ states.
  After interchanging the accepting and rejecting states in dfa $D$,
  we get  dfa $D'$ for  language $L^c$ 
  with the same number of states as in  $D$.
  In  dfa $D'$, all states are accepting, except for state $f$,
  and moreover, accepting state $d$ goes to itself 
  by each symbol since it was dead in $D$.
  The initial state $s$ of  dfa $D'$ is accepting as well.
  Let us construct  nfa $N'$ of $2^{n-1}$ states 
  for $L^c$ from  dfa $D'$ as follows.
  First, add a transition by a symbol $a$ from a state $q$
  to the initial (and accepting) state $s$,
  whenever  there is a transition in  dfa $D'$ from $q$ to $d$ by $a$
  (in particular,  add transition from $d$ to $s$ by each symbol).
  Next,   make state $d$ rejecting. 
  Finally,  redirect all transitions going to state $f$ to state $d$,
  and remove state $f$ with all its ingoing and outgoing transitions. 
  The resulting language is  the same, that is $L^c$,
  and the nfa $N'$ has $2^{n-1}$ states. 
  The prefix-free language $L c$, where $L$ is the binary $(n-1)$-state 
  nfa language   reaching the bound $2^{n-1}$ for complement \cite{ji05},
  meets the bound since  the set
    $\{(x,yc) \mid (x,y)$ is in the fooling set for $L$\  \cite{ji05}$\}$
  is a fooling set of size $2^{n-1}$ for language $L c$.

\label{part4}
  4. The upper bound 
  for intersection of a  prefix-free $m$-state nfa language 
  and a  regular $n$-state nfa language
  is $(m-1)n+1$, and 
  the upper bound for 
  $K\cap L^c$, then follows from part 3.
  For tightness, first let $L'$ be the ternary $n$-state nfa 
  prefix-free language from part 3 meeting the bound $2^{n-1}$ for complement.
  Let $\mathcal{F'}$ be the fooling set for $(L')^c$ described in part 3.
  In each state of the nfa for $L'$, except for final state $n$,
  add a loop by $d$, and denote the resulting prefix-free language by~$L$.
  Next, define an $m$-state nfa prefix-free language $K$
  by $K=((a+b)^*d)^{m-2}(a+b)^*c$.
  Consider the following set 
  $
     \mathcal{F}=\{ (xd\,^i,\,d\,^{m-2-i}y) \mid (x,y)\in \mathcal{F'},
                                                 i=0,1,\ldots,m-2\}.
  $
  For each pair in  $\mathcal{F}$,
  the string $xd\,^id\,^{m-2-i}y$ is in  $K$.
  The nfa for $L$, as well as the nfa for $L'$,
  goes to a subset of $\{1,2,\ldots,n-1\}$ by $x$.
  In each state of this subset, there is a loop by $d$ in the nfa for $L$,
  so the nfa is in the same subset after reading $d^{m-2}$.
  Then it proceeds as the nfa for $L'$ and rejects since $xy$ is in $(L')^c$.
  Thus  $xd\,^id\,^{m-2-i}y\in L^c$. 
  On the other hand, if $i\neq j$,
  then  $xd\,^id\,^{m-2-j}y\notin K$.
  Now assume that $i=j$, and that $(x,y)$ 
  and $(u,v)$ are two distinct pairs in $\mathcal{F'}$.
  Then, without loss of generality, $xv\notin (L')^c$, and so $xv\in L'$.
  Thus there exists an accepting computation of the nfa for $L'$ on  string $xv$.
  It follows that there also exists an accepting computation 
  of the nfa for $L$ on $xv$
  since after reading $x$ the nfa for $L'$ is in a state in $\{1,2,\ldots,n-1\}$,
  in which there is a loop by $d$ in the nfa for $L$.
  Therefore,  $xv\in L$, and so  $xv\notin L^c$.
  Hence  $\mathcal{F}$
  is a fooling set for language $K\cap L^c$ of size $(m-1)2^{n-1}$.
  Now, add one more pair  $(a^{n-2}d^{m-2}c,\eps)$.
  The resulting set is again a fooling set for $K\cap L^c$.
 \hfill$\Box$

 \begin{theorem}[Concatenation, Reversal, Star]\label{thm:con,rev,star}
  Let  $K$ and $L$
  be  prefix-free  languages  with $\nsc(K)=m$ and $\nsc(L)=n$.
  Then \\
  1. $\nsc(KL)\le m+n-1$, and the bound is tight in the unary case; \\
  2. $\nsc(L^R)\le n$, and the bound is tight in the unary case;\\
  3. $\nsc(L^*)\le n$, and the bound is tight in the binary case.
 \end{theorem}

 \noindent\emph{Proof.}
  1. Since both languages are prefix-free,
  to get an nfa for their concatenation,
  we  merge the final state in the nfa for $K$
  with the initial state  in the nfa for $L$.
  This gives the upper bound $m+n-1$.
  For tightness, consider unary  prefix-free regular languages 
  $a^{m-1}$ and $a^{n-1}$.
  Their concatenation  is $a^{m+n-2}$.
  Every singleton language $a^{k-1}$
  is accepted by a $k$-state nfa, and the nfa is minimal
  since  $\{(a^i,a^{k-1-i}) \mid i=0,1,\ldots,k-1\}$
  is a fooling set for such a language.

  2. To obtain an $n$-state nfa for the reversal,
  we reverse all transitions in the nfa for a prefix-free language $L$,
  and switch the role of the initial and the sole accepting state.
  The unary language $a^{n-1}$ meets the bound.

  3. Since  language $L$ is prefix-free,
  we can construct an nfa for  language $L^*$
  from the nfa for $L$, with the initial state $s$, final state $f$,
  and transitions function $\delta$ as follows.
  We make  final  state $f$ initial, thus $\eps$ will be accepted.
  We add transitions by each symbol $a$
  from state $f$  to $\delta(s,a)$.
  The resulting $n$-state nfa recognizes $L^*$.
  For tightness, consider  binary prefix-free language $L=(b^{n-1})^*a$.
  Since the set 
   $\{(\eps,\eps)\}\cup\{(b^i,b^{n-1-i}a) \mid i=1,2,\ldots,n-1\}$
  is a fooling set for  language $L^*$ of size $n$,
  every nfa for the star  requires $n$ states.
 \hfill$\Box$

 \begin{theorem}[Cyclic Shift]\label{thm:cyclic}
  Let $n\ge3$ and let $L$ be a prefix-free regular language  with $\nsc(L)=n$.
  Then $\nsc(L^{cs})\le 2n^2-4n+3$.
  The bound is tight in the binary case.
 \end{theorem}

 \noindent\emph{Proof.}
  The construction in  Theorem~\ref{thm:sc_cyclic}
  gives an $(n-1)(2n-2)$-state nfa for the cyclic shift
  with a set $S$
  of initial states.
  To get an nfa with a single initial state,
  we add a new initial state going to $S$ by the empty string.
  \begin{figure}[b]
  \centerline{\includegraphics[scale=.40]{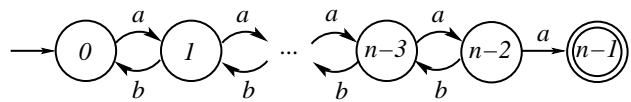}}
  \caption{The prefix-free nfa meeting the bound $2n^2-4n+3$ on cyclic shift.}
  \label{fig:cyclic}
  \end{figure}
  For tightness, consider the binary language
  accepted by the nfa of Figure~\ref{fig:cyclic}.
  The proof proceeds by  describing a fooling set 
  for the cyclic shift of this language
  of size $(n-1)(2n-2)$. Then we use Lemma~\ref{le:fool}
  to prove that one more state is necessary. 
 \hfill$\Box$

\bibliographystyle{eptcs}
\bibliography{bibliography}

\end{document}